\documentclass[%
 reprint,
superscriptaddress,
 amsmath,amssymb,
 aps,
    prl,
]{revtex4-1}

\usepackage{graphicx}
\graphicspath{{figures2/}}
\usepackage{dcolumn}
\usepackage{bm}
\usepackage{hyperref}
\usepackage{textcomp}
\usepackage{xcolor}

\begin{document}

\preprint{APS/123-QED}

\title{A Topology-Controlled Photonic Cavity Based on the Near-Conservation of the Valley Degree of Freedom}

\author{Yandong Li}
 \email{yl2695@cornell.edu}
\affiliation{School of Applied and Engineering Physics, Cornell University, Ithaca, New York 14853, USA}
\author{Yang Yu}
\affiliation{School of Applied and Engineering Physics, Cornell University, Ithaca, New York 14853, USA}
\author{Fengyu Liu}
\affiliation{School of Applied and Engineering Physics, Cornell University, Ithaca, New York 14853, USA}
\affiliation{School of Physics, Nankai University, Tianjin 300071, China}

\author{Baile Zhang}
\affiliation{Division of Physics and Applied Physics, School of Physical and Mathematical Sciences, Nanyang Technological University, 21 Nanyang Link, Singapore 637371, Singapore}
\affiliation{Centre for Disruptive Photonic Technologies, The Photonics Institute, Nanyang Technological University, 50 Nanyang Avenue, Singapore 639798, Singapore}
\author{Gennady Shvets}
 \email{gs656@cornell.edu}
\affiliation{School of Applied and Engineering Physics, Cornell University, Ithaca, New York 14853, USA}

\date{\today}

\begin{abstract}
We demonstrate a novel path to localizing topologically-nontrivial photonic edge modes along their propagation direction. Our approach is based on the near-conservation of the photonic valley degree of freedom associated with valley-polarized edge states. When the edge state is reflected from a judiciously oriented mirror, its optical energy is localized at the mirror surface because of an extended time delay required for valley-index-flipping. The degree of energy localization at the resulting topology-controlled photonic cavity (TCPC) is determined by the valley-flipping time, which is in turn controlled by the geometry of the mirror. Intuitive analytic descriptions of the ``leaky" and closed TCPCs are presented, and two specific designs -- one for the microwave and the other for the optical spectral ranges -- are proposed.
\end{abstract}

\maketitle

The reflection of a photon from a perfect mirror is one of the simplest and most fundamental phenomena in optics. Despite its simplicity, it can be utilized to build one of the most important photonic ingredients -- a Fabry-P\'{e}rot cavity -- when two parallel mirrors are placed along the photon's path and separated by a finite distance $L$. Electromagnetic waves with the wavelenths $\lambda_N$ satisfying a constructive interference condition $L=(N-1/2)\lambda_N$ are trapped in such a cavity by forming a standing wave with $N$ antinodes uniformly spaced between the mirrors. The energy confinement volume of every mode is thus equal to that of a conventional cavity. This picture can be intuitively understood by noting that the photons spend most of their time traveling between the mirrors. In this Letter, we pose, and affirmatively answer, the following question: can we design an optical cavity from a novel perspective of manipulating the reflection time of a photon from the cavity mirror? If a photon can ``stick" to the mirror for a long time $\tau_{\rm refl}$ which is much longer than the time for it to propagate the distance between two adjacent antinodes inside the Fabry-P\'{e}rot cavity, $T_{\rm FP} \sim \lambda_N / \left(2 v_g\right)$, then the optical energy could be expected to strongly localize near the mirror.

The approach presented in this Letter uses photonic topological insulators (PTIs)~\cite{Khanikaev:2012} to engineer the reflection time $\tau_{\rm refl}$. One key feature of topological photonics~\cite{Raghu:2008_1,Raghu:2008_2,ZWang_Marin:2008,ZWang_Marin:2009,Hafezi:2011,Hafezi:2013,Rechtsman:2013,Lu:2014,ShvetsReview2017,OzawaReview2019} is the existence of robust unidirectional edge/kink states propagating either along the PTI edges, or along the interface between two topologically distinct PTIs. Upon encountering photonic defects, such robust modes circumvent them rather than suffer a back-reflection ~\cite{Raghu:2008_1,Raghu:2008_2,ZWang_Marin:2008,ZWang_Marin:2009,YDChong:2008,Hafezi:2011,Hafezi:2013,Rechtsman:2013,FangKJ:2012,Khanikaev:2012,CTChan:2014,Amo:2015,Ma:2015,Ma:2017}. The propagation direction of topologically robust edge/kink (TREK) states is typically linked to a topologically conserved quantity, such as the Chern, spin-Chern, or valley-Chern indices ($C$, $C_s$, and $C_v$, respectively). While $C=0$ for any time-invariant photonic structure, its crystalline symmetries can produce discrete degrees of freedom (DOFs) such as spin and valley ~\cite{Ma:2015,Ma:2016,Ma:2017,Gao:2017,Rechtsman:2018} and endow the structure with non-vanishing topological indices $C_s$ and $C_v$. Therefore, back-scattering a TREK state requires breaking of the bulk topological order of a PTI and changing the corresponding topological index~\cite{Lu:2014,OzawaReview2019}.

In this Letter, we concentrate on the controllable flipping of the valley DOF~\cite{XiaoDi:2007,YaoWang:2008,XiaoDi:2012} achieved through a symmetry-breaking termination of a valley photonic crystal (VPC). The existence and robustness of valley-polarized TREKs has been experimentally demonstrated across the electromagnetic spectrum: from microwave to optical frequencies~\cite{Gao:2017,Litchinitser2019}. A number of novel optical devices based on VPCs have been proposed, including delay lines~\cite{Ma:2016}, quantum optical platforms~\cite{Hafezi2019}, and nanoscale topological waveguides~\cite{Minwoo:2018}. When a valley-polarized TREK is reflected by a terminating mirror due to the reversal of its valley DOF, if the effective reflection time $\tau_{\rm refl}$ is sufficiently long, then the near-conservation of the valley DOF can result in a subwavelength topology-controlled photonic cavity (TCPC) near the mirror. Furthermore, we demonstrate that a non-metallic photonic crystal (PhC) extends this concept to optical frequencies.

\begin{figure}
    \centering
    \includegraphics[width=0.48\textwidth]{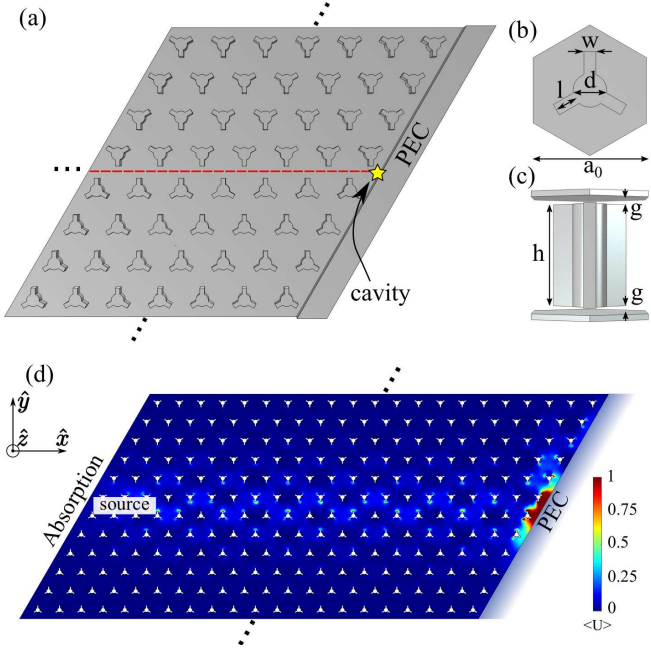}
    \caption{\label{fig:1}(a) A ``leaky" PEC TCPC based on the near-conservation of the valley DOF in a mirror-terminated photonic structure. TREK states are supported by a domain wall (red horizontal line) between two VPCs with opposite $C_v$. (b) Top and (c) side views and geometry definitions of the unit cell. (d) Energy from an excited TREK state is localized at the cavity, for $\omega = 0.75 \left( 2\pi c/a_0 \right)$. Color: time-averaged energy density. Dimensions: $l=0.12 a_0$, $w=0.06 a_0$, $d=0.2 a_0$, $g=0.03a_0$, $h=0.94a_0$. The sketches in (a,b,c) are not to scale.}
\end{figure}

\begin{figure}
    \centering
    \includegraphics[width=0.48\textwidth]{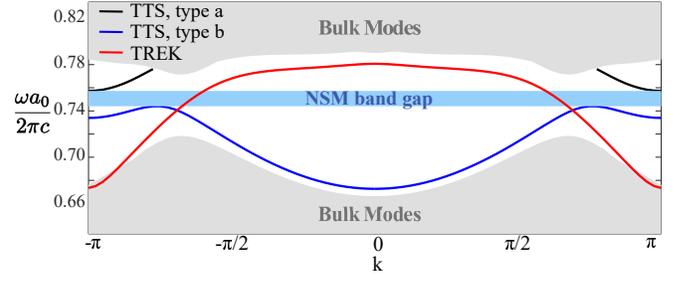}
    \caption{\label{fig:2}Dispersion bands of the TE-polarized TREK (red lines) and TTS (black and blue lines) modes. The wavenumber $k$ is along the domain wall for the TREK and along the PEC mirror for the TTS modes. Black lines: type $a$ TTS mode, propagating downwards (direction $-\hat{x}/2-\sqrt{3}\hat{y}/2$ in Fig.~\ref{fig:1}(d)) along the mirror; Blue line: type $b$ TTS mode, propagating upwards (opposite to type $a$) along the mirror. Only TREK modes can propagate inside the NSM bandgap. Structure parameters: same as in Fig.~\ref{fig:1}.}
\end{figure}

The first example of a TCPC is a structure comprised of a triangular lattice of tripod-shaped  perfect electric conducting (PEC) pillars~\cite{Ma:2016,Ma:2017} symmetrically inserted between two parallel plates separated by the distance $h_0$ (shown in Figs.~\ref{fig:1}(a,b,c)). The structure contains a domain wall separating two topologically-nontrivial VPCs that are mirror-reflections of each other about the $x-z$ plane (where the $x-$axis is chosen along the nearest-neighbor direction of the lattice). Therefore, the two VPCs have opposite valley-Chern indices $C_v=\pm 1/2$~\cite{Ma:2016}, and the resulting domain wall supports $K^{\prime}$ ($K$)-valley-polarized TREK states propagating along the forward (backward) directions~\cite{Ma:2016,Ma:2017,Gao:2017}. The VPC supports electromagnetic modes whose in-plane electric field components $E_{x,y}$ are either symmetric (TE modes) or anti-symmetric (TM modes) with respect to the $z=h_0/2$ mirror symmetry plane~\cite{PhC}. In what follows, we focus on the TE TREK modes whose dispersion band $\omega_{\rm TREK}(k)$ is plotted in Fig.~\ref{fig:2} as a red line.

The structure is terminated by a PEC mirror producing inter-valley scattering at a rate $\tau_{\rm refl}^{-1}$ which, we assume, is determined by its shape and orientation. As an example, we consider a straight mirror oriented at $60^{\circ}$ with respect to the propagation direction of the $K$-valley ($k\sim 4\pi/3a$) TREK mode. When a forward-propagating TREK state encounters the mirror, it has three distinct scattering channels.
The first and the second channels involve scattering either up or down the mirror/VPC interface into topologically-trivial surface (TTS) modes.
The dispersion curves of the TTS modes are plotted in Fig.~\ref{fig:2} as black (blue) lines corresponding to downward (upward) modes labeled as type $a$($b$). The TTS modes are localized near the terminating mirror (see Supplemental Material). Because the TTS modes do not span the entire bandgap, there exists a frequency range labelled as a no-surface-mode (NSM) bandgap shown in Fig.~\ref{fig:2}.
When a TE-polarized TREK mode is inside the NSM band bandgap, it cannot scatter into either of the two TTS modes.
The third scattering mechanism is backscattering, and it requires the reversal of the mode's valley index from $K$ to $K^{\prime}$. Therefore, valley-flipping rate is identical to the decay rate of the ``leaky" VPC-based TCPC.

Previous studies revealed that the rate of valley-flipping depends on the geometry of the perturbations ~\cite{Ma:2016,Gao:2017,Rechtsman:2018}. For example, perturbations along the principal axes of the lattice (the so-called zigzag directions) significantly suppress inter-valley scattering (see Supplemental Material). On the other hand, perturbations along the orthogonal (armchair) directions produce much higher rates of inter-valley scattering. It is expected that the zigzag orientation of the PEC mirror chosen in Fig.~\ref{fig:1}(a) should minimize (albeit not entirely eliminate) valley-flipping. The evanescent nature of the TREK states along the direction normal to that of their propagation is responsible for the residual rate of valley-flipping (see Supplemental Material for the detailed calculation).

Within the NSM bandgap, this ``leaky" TCPC can be analytically described as a resonator at an abruptly terminated end~\cite{Mann:19,HassaniGangaraj:19,Buddhiraju:2018,Tsakmakidis:2017} of a topologically robust port. The driven Lorentzian response of the resonator to a time-harmonic input signal with frequency $ \omega \equiv  \omega_0 + \Delta \omega $ is:
\begin{equation}
|A|^2 = |s^+|^2 \frac{2\gamma}{(\omega-\omega_0)^2 + \gamma^2},
\label{equ:Lorentzian}
\end{equation}
where $A$ and $s^+$ are the complex-valued mode amplitude and incoming wave amplitude, $\omega_0$ and $\gamma$ are the resonator eigenfrequency and decay rate (see Supplemental Material for details).
We extract $\omega_0$ and $\gamma \equiv \tau_{\rm refl}^{-1}$ of the VPC-based TCPC shown in Fig.~\ref{fig:1} by numerically fitting the simulated energy content inside the cavity as a function of $\Delta \omega$ (see Supplemental Material).
The extracted cavity parameters are $\omega_0 \approx 0.75 \left( 2\pi c/a_0\right)$ and $\gamma \approx 3.1\times 10^{-3} \left( 2\pi c/a_0\right)$. The reflection time is hence $\tau_{\rm refl} \approx 51.3 \left(a_0/c\right)$.

Note that $\tau_{\rm refl} \gg T_{\rm FP} \sim a_0/v_g \approx 2.9 \left( a_0/c \right)$, where $a_0$ is the lattice constant and $v_g$ is the group velocity of the TREK mode.
This result confirms our initial conjecture that a terminating mirror of the zigzag type indeed presents a weak valley-flipping perturbation to a propagating TREK state, delaying the reflection and resulting in energy confinement.
The ``leaky" resonator with a $Q$-factor equal to $Q \equiv \omega_0/ \left( 2\gamma \right) \approx 121$ is based on the near-conservation of the valley DOF.

\begin{figure}
    \centering
    \includegraphics[width=0.48\textwidth]{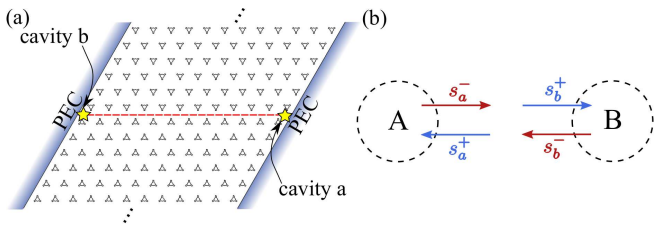}
    \caption{\label{fig:3}(a) A valley-based TCPC terminated by PEC reflectors on both ends. (b) The corresponding one-port-two-resonators model of a TCPC.}
\end{figure}

The near-conservation of the valley DOF can also be utilized for designing closed Fabry-P\'{e}rot resonators.
An example of such a cavity terminated by PEC mirrors on both ends of the domain wall is shown in Fig.~\ref{fig:3}.
Below we demonstrate that the near-conservation of the valley DOF produces electromagnetic mode structures that can be very distinct from those in conventional Fabry-P\'{e}rot resonators.
Specifically, if $\tau_{\rm refl} \gg T_{\rm FP}$ is satisfied, then the modes are strongly confined at surfaces of the terminating mirrors.
On the other hand, when this condition is violated, the modes are uniformly distributed between the mirrors as expected for a Fabry-P\'{e}rot resonator.

To describe the closed TCPC, we develop an analytic one-port-two-cavity model schematically illustrated in Fig.~\ref{fig:3}(b) and described by the following coupled differential equations for the complex-valued cavity mode amplitudes $A$ and $B$~\cite{Haus,CMT_SHFan:2003,CMT_SHFan_2004,PhC}:
\begin{equation}
\begin{split}
\frac{dA}{dt}=-(i\omega_0+\gamma)A+K_a s_a^+,\\
\frac{dB}{dt}=-(i\omega_0+\gamma)B+K_b s_b^+,
\end{split}
\label{cmt:1}
\end{equation}
where the incoming (outgoing) wave amplitude $s_{a(b)}^{+(-)}$ coupled to cavity $a$($b$) are calculated as
\begin{equation}
s_a^- = C_a s_a^+ +D_a A \ \ {\rm and} \ \ s_b^- = C_b s_b^+ +D_b B.
\label{cmt:2}
\end{equation}
Here $K_{a,b}$ and $D_{a,b}$ are the coupling-in and -out coefficients of cavities $a$ and $b$, respectively, and $C_{a,b}$ are the corresponding scattering coefficients.
From the symmetric orientation of the two terminating mirrors, we further conclude that the two cavities are identical and, therefore, have the same eigenfrequency $\omega_0$ and decay rate $\gamma$, as well as identical scattering and coupling coefficients: $D_a=D_b \equiv D$, $K_a=K_b \equiv K$, and $C_a=C_b \equiv C$.
In the absence of non-radiative losses, energy conservation and time-reversal symmetry impose the following constraints~\cite{CMT_SHFan:2003,CMT_SHFan_2004}: $D^{\dagger}D = 2\gamma$, $K=D$, and $C D^* = -D$.
Without loss of generality, we set $K=D=\sqrt{2 \gamma}$ and $C=-1$.
The only remaining relation is between the outgoing wave from cavity $b$($a$) and the incoming wave into cavity $a$($b$).
Accounting for the phase shift $\alpha = -2\pi\frac{L}{\lambda}$ accrued along the path of the TREK modes, the following relations are obtained: $s_a^+=e^{i\alpha}s_b^-$ and $s_b^+=e^{i\alpha}s_a^-$, where the negative sign of $\alpha$ is due to the negative index of the TREK modes.

The eigenfrequencies of the symmetric and anti-symmetric eigenmodes of the cavity can be found analytically (see Supplemental Material for modes classification and analytic details):
\begin{equation}
\omega_{\rm sy}=\omega_0-\gamma\tan{(\alpha/2)} \ \ {\rm and} \ \
\omega_{\rm an}=\omega_0+\gamma\cot{(\alpha/2)},
\label{cmt:omega}
\end{equation}
where $\omega_{\rm sy \left( an \right)}$ are the eigenfrequencies of the symmetric (anti-symmetric) eigenmodes.

This analytic model reveals a characteristic feature of the symmetric and anti-symmetric modes: the ratio of the energy inside the cavities, $U_{\rm cav}\propto \left( |A^2| + |B^2| \right)$, to the energy per antinode inside the topological waveguide, $U_{\rm FP}\propto \sum_{i=a,b} |s_i|^2 a_0/v_g $, is inversely proportional to the decay rate $\gamma$:
\begin{equation}\label{eq:concentration}
  \frac{U_{\rm cav}^{\rm sy}}{U_{\rm FP}} \propto \frac{2 \cos^2(\alpha/2)}{\gamma a_0/v_g} \ \ {\rm and} \ \ \frac{U_{\rm cav}^{\rm an}}{U_{\rm FP}} \propto \frac{2 \sin^2(\alpha/2)}{\gamma a_0/v_g}.
\end{equation}
Our model predicts that the energy is more localized in the cavities when the decay rate $\gamma$ is smaller (see Figs.~\ref{fig:fitting_comsol_cmt}(c,d) and Supplemental Material for details).

\begin{figure}
    \centering
    \includegraphics[width=0.48\textwidth]{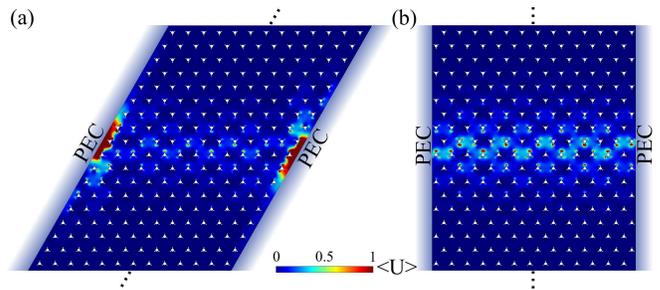}
    \caption{\label{fig:comsol_zig_arm} Examples of the time-averaged energy distributions of the eigenmodes of a TCPC. (a) With zigzag terminations, energy is mostly localized in the two cavities. (b) With armchair terminations, energy is distributed in the waveguide. Horizontal cavity length is $L=13a_0$ for both cases.}
\end{figure}

\begin{figure}
    \centering
    \includegraphics[width=0.48\textwidth]{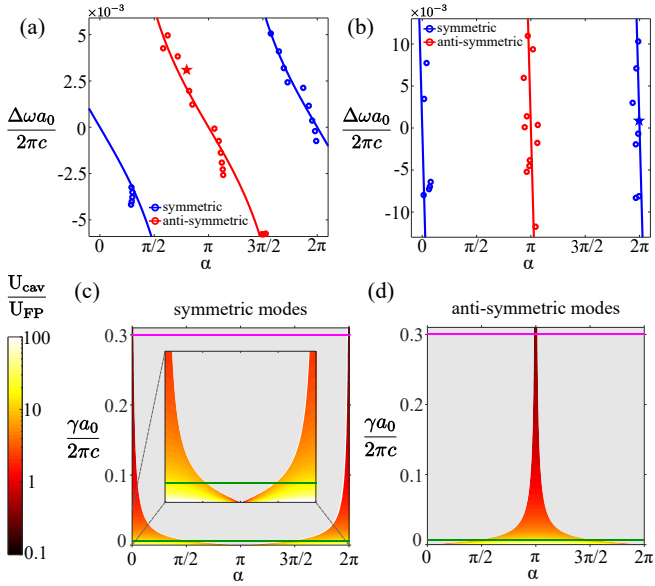}
    \caption{\label{fig:fitting_comsol_cmt}(a,b) Symmetric and anti-symmetric eigenmodes data of the TCPC with (a) zigzag terminations and (b) armchair terminations. The circles are the numerical results from COMSOL; the lines are the analytic results from the coupled mode theory; the data points labelled by stars correspond to the field profiles shown in Fig.~\ref{fig:comsol_zig_arm}. (c,d) The ratio between the energy stored in the two cavities and the topological waveguide for (c) symmetric modes and (d) anti-symmetric modes. Green lines represent $\gamma_{\rm zig} a_0 / \left( 2\pi c \right)$; magenta lines represent $\gamma_{\rm arm} a_0 / \left( 2\pi c \right)$. Only the solutions close to the center of the NSM bandgap are plotted; i.e., $\omega_0 - \Delta\omega < \omega < \omega_0 + \Delta\omega$, are plotted, where $\omega_0 = 0.75 \left( 2\pi c/a_0 \right)$ and $\Delta\omega = 0.01\omega_0$. The grey region represents the combinations of $\gamma$ and $\alpha$ corresponding to frequencies outside of that range, determined by Eq.~\eqref{cmt:omega}.}
\end{figure}

As concluded previously, the decay rate from the cavity is the rate of valley-flipping at the PEC boundary, which strongly depends on the type of geometry (zigzag or armchair) of the VPC termination. Therefore, we expect that the energy is mostly localized in the cavities with zigzag terminations, whereas, with armchair terminations, the energy is mostly stored inside the waveguide. This intuitive conclusion is confirmed by the result of COMSOL simulations of the cavity eigenmodes inside the two types of platforms shown in Fig.~\ref{fig:comsol_zig_arm}: the zigzag-oriented mirrors in Fig.~\ref{fig:comsol_zig_arm}(a) indeed produce much stronger field localization than the armchair-oriented mirrors in Fig.~\ref{fig:comsol_zig_arm}(b).

Next, we check the applicability of the analytic model to the realistic system shown in Fig.~\ref{fig:comsol_zig_arm}. By varying the path length of the TREK state from $L_{\rm min}=6a_0$ to $L_{\rm max} = 21a_0$ (with $\Delta L = a_0$ per increment), we vary the phase delay $\alpha$. For each length $L_{\rm min} \leq L \leq L_{\rm max}$ of the topological cavity simulated using COMSOL, we extract the frequencies $\omega_{\rm sy,an}$ of the symmetric and anti-symmetric eigenmodes and calculate $\gamma$ by fitting the simulation results to Eq.~\eqref{cmt:omega} (Fig.~\ref{fig:fitting_comsol_cmt}).
We have found that $\gamma_{\rm zig} \approx 6.4 \times 10^{-3} \left( 2\pi c/a_0\right)$ for the zigzag termination,  and $\gamma_{\rm arm} \approx 3.0\times 10^{-1} \left( 2\pi c/a_0\right)$ for the armchair termination.

Remarkably, the zigzag termination delays the reflection for an extra time of $\tau = 1/\gamma_{\rm zig} - 1/\gamma_{\rm arm} \approx 24.4 \left( a_0/c \right)$ as compared to the armchair termination. Such long delay is consistent with the high $Q$ value of the leaky single-mirror TCPC described earlier. It is also in agreement with the time delay calculated by numerically processing the back-reflected wave of the COMSOL simulation (see Supplemental Material). Moreover, it is possible to increase the reflection time by further suppressing the inter-valley scattering at the PEC surface, through optimizing the geometrical detail of the zigzag termination (see Supplemental Material).
We observe the following hierarchy of the time scales: $\tau_{\rm refl,zig} \gg T_{\rm FP} \gg \tau_{\rm refl,arm}$.
Also, for the armchair termination, the possible phase shifts $\alpha$ are restricted to $\alpha \approx 2\pi N$ for symmetric or $\alpha \approx 2\pi (N+1/2)$ for anti-symmetric modes ($N$ is an integer number). These conclusions are confirmed by Fig.~\ref{fig:fitting_comsol_cmt}(b). The large valley-flipping rate at the armchair termination makes the reflection happens almost instantaneously, thereby restoring all conventional properties of a Fabry-P\'erot cavity.

\begin{figure}
    \centering
    \includegraphics[width=0.48\textwidth]{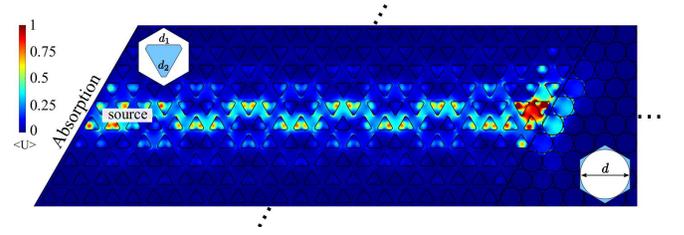}
    \caption{\label{fig:8} An all-dielectric TCPC. Energy  from  an  excited TREK state is localized at the cavity. The VPC region is formed by triangular Si rods; the trivial PhC region is a Si slab with arrayed holes. Color: time-averaged energy density. Frequency corresponds to $\omega = 0.43 \left( 2\pi c /a_0 \right)$. $d_1=0.64a_0$, $d_2=0.09a_0$, $d=0.98a_0$, and $\epsilon_{\rm Si}=13$.}
\end{figure}

Furthermore, we demonstrate that the TCPC concept can be readily extended to optical frequencies by using all-dielectric VPCs~\cite{Ma:2016}, and replacing metal mirrors with their PhC counterparts. The dispersion of the TTS along the VPC/PhC interface (see Supplemental Material) defines the bandgap inside which no energy can leak through the interface. Therefore, when a TREK state encounters the interface, its reflection is delayed by the valley-flipping time, and its optical energy is concentrated at the junction between the two VPCs and the terminating trivial PhC (Fig.~\ref{fig:8}). Recent experimental developments in fabricating and characterizing all-dielectric topological VPCs~\cite{Litchinitser2019,Hafezi2019,Singh2019} make us optimistic about near-future prospects for realizing optical TCPCs and using them for a variety of quantum optics applications~\cite{Lukin:2018,Segev:2018,Waks:2018}.

In conclusion, we have demonstrated the localization of topologically protected TREK based on the near-conservation of the valley DOF.
This phenomenon can be used for making compact topology-controlled photonic resonators, and no time-reversal-symmetry-breaking is required.
Compared to the nanoplasmonic localization achieved through adiabatic tapering~\cite{adiabatictaper:2004}, this novel mechanism does not need an extended tapering region. We also demonstrate a nano-fabricable all-dielectric realization of the TCPCs using a combination of topologically trivial and non-trivial photonic crystals.
We expect that the TCPC concept will benefit the development of nonlinear and quantum optical devices and technologies.
\begin{acknowledgments}
This work was supported by the Office of Naval Research (ONR) under a Grant No.~N00014-17-1-2161, by the National Science Foundation (NSF) under a Grant No.~DMR-1741788, and by the Cornell Center for Materials Research with funding from the NSF MRSEC program (DMR-1719875). Y.L. was supported in part by Cornell Graduate School Fellowship. F.L. acknowledges financial support from the Pilot Scheme of Talent Training in Basic Sciences (Boling Class of Physics, Nankai University). The authors would like to thank helpful discussions with Minwoo Jung, Haoran Xue, and Ran Gladstone.

\end{acknowledgments}

\bibliographystyle{apsrev4-1}

\end{document}